\def\figdir{.}

\documentclass[english,prb,longbibliography,notitlepage,preprintnumbers]{revtex4-1}

\usepackage[T1]{fontenc}
\usepackage[latin9]{inputenc}
\usepackage{bm}
\usepackage{amsmath}
\usepackage{graphicx}
\usepackage{babel}

\usepackage{datetime2} 
\usepackage{currfile} 
\DTMsetstyle{default}\DTMsettimestyle{hmmss}
\begin{document}

\global\long\def\Cm{C_{{\scriptscriptstyle M}}}

\global\long\def\va{\bm{a}}
\global\long\def\vb{\bm{b}}
\global\long\def\vj{\bm{j}}
\global\long\def\vk{\bm{k}}
\global\long\def\vn{\bm{n}}
\global\long\def\vp{\bm{p}}
\global\long\def\vq{\bm{q}}
\global\long\def\vr{\bm{r}}
\global\long\def\vs{\bm{s}}
\global\long\def\vu{\bm{u}}
\global\long\def\vv{\bm{v}}
\global\long\def\vw{\bm{w}}
\global\long\def\vx{\bm{x}}
\global\long\def\vy{\bm{y}}
\global\long\def\vz{\bm{z}}

\global\long\def\vA{\bm{A}}
\global\long\def\vB{\bm{B}}
\global\long\def\vD{\bm{D}}
\global\long\def\vE{\bm{E}}
\global\long\def\vF{\bm{F}}
\global\long\def\vH{\bm{H}}
\global\long\def\vJ{\bm{J}}
\global\long\def\vK{\bm{K}}
\global\long\def\vL{\bm{L}}
\global\long\def\vN{\bm{N}}
\global\long\def\vP{\bm{P}}
\global\long\def\vR{\bm{R}}
\global\long\def\vS{\bm{S}}

\global\long\def\val{\bm{\alpha}}
\global\long\def\vom{\bm{\omega}}
\global\long\def\vga{\bm{\gamma}}
\global\long\def\vep{\bm{\epsilon}}
\global\long\def\vnabla{\bm{\nabla}}
\global\long\def\vmu{\bm{\mu}}
\global\long\def\vnu{\bm{\nu}}
\global\long\def\vsi{\bm{\sigma}}
\global\long\def\vSi{\bm{\Sigma}}

\global\long\def\order#1{\mathcal{O}\left(#1\right)}

\global\long\def\edge#1{\left.#1\right|}
\global\long\def\d{\mbox{d}}

\preprint{Alberta Thy 9-18}
\title{Acceleration due to buoyancy and mass renormalization}

\author{Kyle McKee}

\affiliation{Department of Physics, University of Alberta, Edmonton, Alberta,
Canada T6G 2E1}

\author{Andrzej Czarnecki}

\affiliation{Department of Physics, University of Alberta, Edmonton, Alberta,
Canada T6G 2E1}
\begin{abstract}
  The acceleration of a light buoyant object in a fluid is analyzed.
  Misconceptions about the magnitude of that acceleration are briefly
  described and refuted. The notion of the added mass is explained and
  the added mass is computed for an ellipsoid of revolution. A simple
  approximation scheme is employed to derive the added mass of a
  slender body. The slender-body limit is non-analytic, indicating a
  singular character of the perturbation due to the thickness of the
  body.  An experimental determination of the acceleration is
  presented and found to agree well with the theoretical
  prediction. The added mass illustrates the concept of mass
  renormalization in an accessible manner.
\end{abstract}
\maketitle
\section{Introduction}

Imagine holding a piece of cork under water. When released, it surges towards
the surface. What is its acceleration? This should be an easy question,
at least when the cork is still moving slowly so that drag can be
neglected. Surprisingly, textbooks and practicing physicists express
a spectrum of conflicting opinions. 

Some physics textbooks\citep{YF14,meirovitch2010fundamentals} suggest
that the mass of the body $m_{b}$ alone determines its acceleration
$a$ due to the balance of the forces of buoyancy $F_{B}$ and weight
$m_{b}g$. This suggests an arbitrarily large acceleration if the
water density greatly exceeds the density of the cork. For a partially submerged cork
on the surface, very large frequency of small oscillations follows,
too. 

Another point of view accounts for the water that has to move
to make room for the accelerating cork: as the cork proceeds upwards,
an equal volume of water accelerates downwards. The two accelerations
have opposite signs but equal magnitudes. It is tempting to conclude
that this magnitude cannot exceed the standard gravitational acceleration
$g$, since ``the maximum acceleration for the displaced fluid is
the free fall acceleration.''\citep{physicsmyths} 
Although this statement is not completely accurate, accounting for the
fluid's acceleration motivates the concept of \emph{added mass}:
in Newton's law
$a=\left(F_{B}-m_{b}g\right)/m$, the mass of the displaced fluid
$m_{f}$ supplements the cork mass  in some measure,
\begin{equation}
m=m_{b}+\Cm m_{f},\label{Cm}
\end{equation}
with a dimensionless added-mass coefficient $\Cm$. George Green first
introduced the concept of added mass, in the context of submerged
pendulums, about two centuries ago.\citep{GeorgeGreenPapers} 

Added mass in a fluid is analogous to the mass renormalization of
subatomic particles, as pointed out by Sydney Coleman in his famous,
yet unpublished, quantum field theory lectures.\citep{Connes-Motives}
The increased inertia of a body immersed in fluid is similar to the
charged particle mass modified by an interaction with a gauge
field. Moreover, the subatomic particle mass renormalization depends
on its surroundings, be it perfect vacuum, a cavity, or an atomic
bound state. Similarly, the added mass changes in the proximity of a
wall or of the fluid surface. Both phenomena are non-dissipative.

This analogy is pedagogically valuable: it is not trivial to develop
an intuition for the mass generation of a subatomic particle, due to
the Higgs field and the charge-selfinteraction.  The elementary
particle context is shrouded in mathematical intricacies, sometimes
including divergent integrals.  It easier to explain how a body can be
weighed outside a fluid and how its inertia is modified in a fluid. In
fact, even a divergent integral will be encountered when 
some degrees of freedom are ignored (see Section \ref{sec:Simple-derivation}).

For a sphere, the added mass amounts to half of the mass of the fluid
it displaces (see Appendix \ref{sec:Added-mass-sphere}), as has been
demonstrated in a beautiful recent
experiment.\citep{messer:2010aa,pantaleone:2011aa} This value of
$\Cm=\frac{1}{2}$ leads to the maximum surge acceleration of twice
$g$, if the mass of the displaced fluid far exceeds the body mass, as
in the case of a ping-pong ball under water. This leads to a third
misconception, namely that $2g$, following from the buoyancy force
acting both on the body and on the fluid it displaces, is the maximum
acceleration of a surging body, independent of its shape.

The present paper is intended to popularize the notion of the added
mass. A theoretical argument in Section \ref{sec:Simple-derivation}
shows that the acceleration can significantly exceed $2g$, and that
acceleration does indeed depend on the geometry of the body and on its
orientation. In addition to a detailed exact calculation for a prolate
ellipsoid (Appendix \ref{sec:Added-mass-ellipsoid}), the simple
calculation in Section \ref{sec:Simple-derivation} is found to
reproduce the leading added-mass effect of a slender body. An
experimental proof of high acceleration is presented in Section
\ref{sec:Experiment}, and may be easily reproduced in a classroom or
on a field trip to a lake.

\section{Simple derivation of the added mass \label{sec:Simple-derivation}}

When a slender body is moving with velocity $\vv$ with respect to a fluid,
the fluid parts to make room for it, then closes behind the body.
The resulting fluid motion is predominantly perpendicular to $\vv$
(transversal).
We denote the fluid's kinetic energy by $T$.

Consider an elongated (prolate) ellipsoid of revolution with the long
semi-axis $c$ and two equal short semi-axes $b<c$. Define the slenderness
parameter $\epsilon=b/c$. If the body is slender, $\epsilon\ll1$,
and moving along $c$, $T$ is small in the sense that it vanishes
with $\epsilon$: when the body is collapsed into a line, its motion
does not disturb the fluid. However, this limit will turn out to be
nontrivial, involving both a power and a logarithm of $\epsilon$.
The thickness of a slender body is an example of a singular perturbation.\citep{historySingPert}

\begin{figure}[h]
\centering\includegraphics[width=0.6\columnwidth]{\figdir/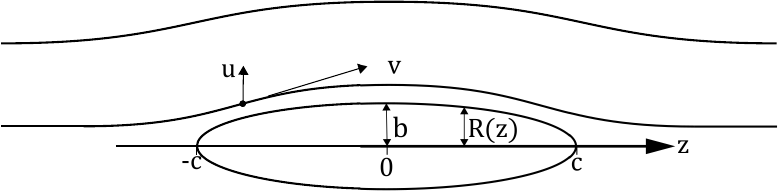}

\caption{Velocity distribution of the fluid around a slender body. The streamline
at which the velocity is indicated is supposed to be very close to
the body. The transverse component $u$ of velocity is related to the total
speed by the slope $R^{\prime}\left(z\right)$ of the body cross-section
at the corresponding value of the coordinate $z$.\label{fig:Velocity-distribution}}
\end{figure}

In order to determine $T$, consider the velocity distribution of
the fluid shown in Figure \ref{fig:Velocity-distribution}. The long
axis of the ellipsoid lies on the $z$-axis and the origin of the
coordinate system is at the center of the ellipsoid. Transverse directions
are parametrized by the azimuthal angle $\varphi$ and the distance
$r$ from the $z$-axis. The boundary of the ellipsoid is axisymmetric
and described by $r=R\left(z\right)$, 
\begin{equation}
\left(\frac{z}{c}\right)^{2}+\left(\frac{R\left(z\right)}{b}\right)^{2}=1.\label{ellips:boundary}
\end{equation}

The sloped surface of the body accelerates the fluid in the transverse
direction. Far from the body, the fluid travels also longitudinally
and returns towards the $z$-axis in the downstream half of the body.
In the rest frame of the fluid far from the body, the streamlines
form a dipole-like pattern. Since the velocity far from the body is insensitive to the slenderness $\epsilon$,
the far-field contribution of the fluid is neglected to uncover the
leading logarithmic dependence of added mass on $\epsilon$. 

In the slender-body approximation, motion in each plane
$z=\text{ constant}$ is treated independently from other planes. This
approximation originates with the 1924 work on Zeppelin
airships.\citep{munk:1924aa} The
transverse velocity at any $r$ is thus related to its value at the boundary
$R\left(z\right)$ by the continuity equation
$2\pi ru\left(r,z\right)=2\pi
R\left(z\right)u\left(R\left(z\right),z\right)$,
\begin{equation}
u\left(r,z\right)=\frac{R\left(z\right)u\left(R\left(z\right),z\right)}{r}.
\end{equation}
The value at the boundary follows from the slope of the surface (see
Fig.~\ref{fig:Velocity-distribution}). Within the slender-body
approximation, the angle between the velocity of the fluid $\vv$ and the
axis of the body is small and its sine and tangent are approximately
equal, thus
\begin{equation}
u\left(R\left(z\right),z\right)=R^{\prime}\left(z\right)v,
\end{equation}
so that finally 
\begin{equation}
u\left(r,z\right)=\frac{R\left(z\right)R^{\prime}\left(z\right)}{r}v.
\end{equation}
This is sufficient to determine the fluid's kinetic energy $T$,
\begin{equation}
T=\frac{\rho}{2}\int_{\text{fluid}}\d^{3}ru^{2}\left(r,z\right)
=\pi\rho v^{2}\int_{-c}^{c}\left(RR^{\prime}\right)^{2}\d z
\int_{R\left(z\right)}^{\infty}\frac{\d r}{r}.
\label{axisymT}
\end{equation}
The divergence of the $r$-integration at large distances is an
artifact of the approximation that the motion in each plane
$z = \text{constant}$ is independent of the motion in other planes. In
fact, of course, far from the axis there is some longitudinal motion
of the fluid and the radial velocity decreases faster than $1/r$. This
happens at distances of the order of the length of the moving body.

However, for the purpose of determining the coefficient of the leading
logarithm (of the slenderness parameter), that large-distance behavior
need not be known. It is sufficient to know the logarithmic
contribution at the lower limit. We replace the upper limit of
the $r$-integration in eq.~\eqref{axisymT} by a cutoff $\Lambda$ of
the order of the length of the body because the velocity field outside
of this falls off faster than $1/r$, and so does not contribute to the logarithmic
singularity in $\epsilon$.

Eq.~(\ref{axisymT}) is valid for any axisymmetric slender body.
For an ellipsoid it gives
\begin{align}
R\left(z\right) & =b\sqrt{1-\left(\frac{z}{c}\right)^{2}}\\
T & \overset{\epsilon\to0}{\sim}\pi\rho v^{2}
\int_{-c}^{c}\left(RR^{\prime}\right)^{2}\ln\frac{\Lambda}{R\left(z\right)}\d z\\
 & = \frac{2\pi}{3}\rho v^{2}cb^{2}\epsilon^{2}\left( \ln\frac{1}{\epsilon} + \order{1}\right).
\end{align}
The volume of the ellipsoid being $\frac{4}{3}\pi cb^{2}$, within
logarithmic accuracy this result
becomes 
\begin{equation}
T\sim\frac{m_{f}v^{2}}{2}\epsilon^{2} \ln\frac{1}{\epsilon},\label{eq:T}
\end{equation}
where $m_{f}$ is the mass of the fluid displaced by the ellipsoid.
The total kinetic energy of the body and the fluid is
\begin{equation}
T=\left(m_{b}+\Cm m_{f}\right)\frac{v^{2}}{2}\label{eq:KE}
\end{equation}
where $m_{b}$ is the mass of the body. Comparison of Eqs.~(\ref{eq:T})
and (\ref{eq:KE}) gives the added mass coefficient (cf.~Eq.~\ref{Cm}),
\begin{equation}
\Cm\overset{\epsilon\to0}{\sim}\epsilon^{2}\ln\frac{1}{\epsilon}.
\end{equation}
For $\epsilon\simeq0.3$, this gives $\Cm\simeq 0.1$. A body of such shape could
reach an acceleration of up to about $\frac{g}{\Cm}\simeq10g$. This
leading order behaviour of the added mass coefficient is consistent
with both the analytic solution in Appendix
\ref{sec:Added-mass-ellipsoid} and earlier slender body
studies.\citep{Handlesman} The following section describes
experimental evidence of an acceleration exceeding $6g$.

\section{Experiment \label{sec:Experiment}}

\subsection{Setup}

\begin{figure}[h]
\begin{centering}
\includegraphics[width=0.35\columnwidth]{\figdir/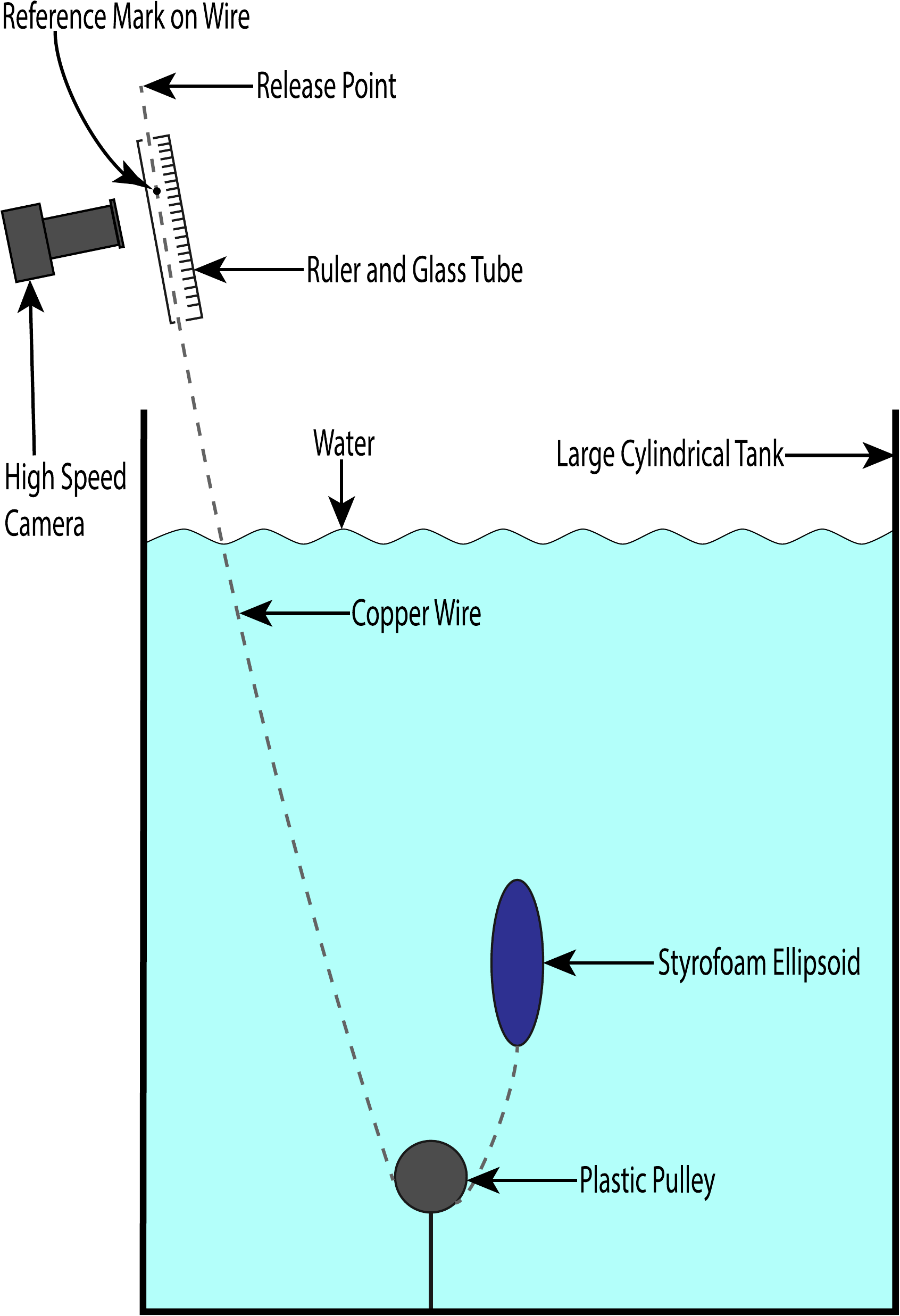} 
\par\end{centering}
\caption{The experimental setup. A thin copper wire is attached to the bottom
of the approximate ellipsoid (spindle) made of styrofoam. The wire
is fed through a pulley, then through a glass tube placed against
a ruler. The motion of a reference mark on the wire is registered
with a high speed camera. \label{fig:The-experimental-setup.}}
\end{figure}

The experimental setup is shown in Fig.~\ref{fig:The-experimental-setup.}. 
A spindle-shaped piece of styrofoam approximates an ellipsoid. Thin 
(0.005 inch or about 0.1 mm diameter) copper wire is attached to the bottom 
end of the spindle, which is completely submerged in a large
cylindrical tank of water. The wire passes through a light plastic
pulley at the bottom of the tank, and its end is held
above the water surface. 

A white mark on the wire facilitates tracking of the wire's motion
once it is released and pulled by the accelerating spindle. The
section of the wire containing the white mark passes through a glass
tube that guides the wire along a millimeter precision ruler in front
of a high-speed camera.  The field of view of the camera includes
about 12 cm of the distance the white mark travels.  The advantage of
this arrangement is that all photography and measurements are made
above water.

A perfect material for producing the spindle is the so-called extruded
polystyrene foam. Produced by Dow Chemical Company and marketed as
Styrofoam Cladmate CM20, it is a readily available and inexpensive
construction insulation material, able to withstand compression. It is
easily cut with hot wire and finely shaped with abrasive paper on a
stationary belt sander. Its density is only about 1/40 of water and
its moisture-resistance prevents water from penetrating, protecting
its low inertia. The spindle is 28 cm long and has about 6 cm mean
diameter in its thickest cross section. A spindle volume of 0.44 liter
is determined from its mass, 10.8 grams. The spindle shape differs
from an ellipsoid by protruding tips. Its slenderness parameter is
thus estimated by $\epsilon\simeq0.3$, assuming the length of an
equivalent ellipsoid to be about 20 cm.

In order to attach the copper wire to the spindle, the wire is inserted
in a small incision of about one inch depth and secured with cyanoacrylate Super
Glue. The thin copper wire is prone to twisting, breaking, and falling
off the pulley, and takes some practice to work with. However, it
has a much higher Young modulus than a fishing line that was initially
employed. The wire tension due to the buoyancy of the spindle is about
4 N, at which the 1.5-meter wire stretches less than a centimeter,
much less than the distance over which the motion is observed. This
removes the only systematic uncertainty that was identified to possibly
\emph{increase } the acceleration of the wire mark compared to the
actual acceleration of the spindle. 

All other effects: drag, friction, finite size of the tank, mass of
the wire and the moment of inertia of the pulley act to decrease the
observed acceleration. Since the goal was to prove that the acceleration
can exceed $2g$ rather than to precisely measure it, these systematic effects
are unimportant for our purposes.

The high-speed camera is NAC Image Technology's HotShot HS1280cc,
set to 700 frames per second. At this speed, very strong light is
required, so the marked part of the wire is illuminated with a Lowel
Pro-Light with a 250 W halogen bulb by Osram. To eliminate ambient
light, a sheet of dark construction paper is used as a backdrop. 

\subsection{Results}

\begin{figure}[h]
\centering\includegraphics[scale=0.5]{\figdir/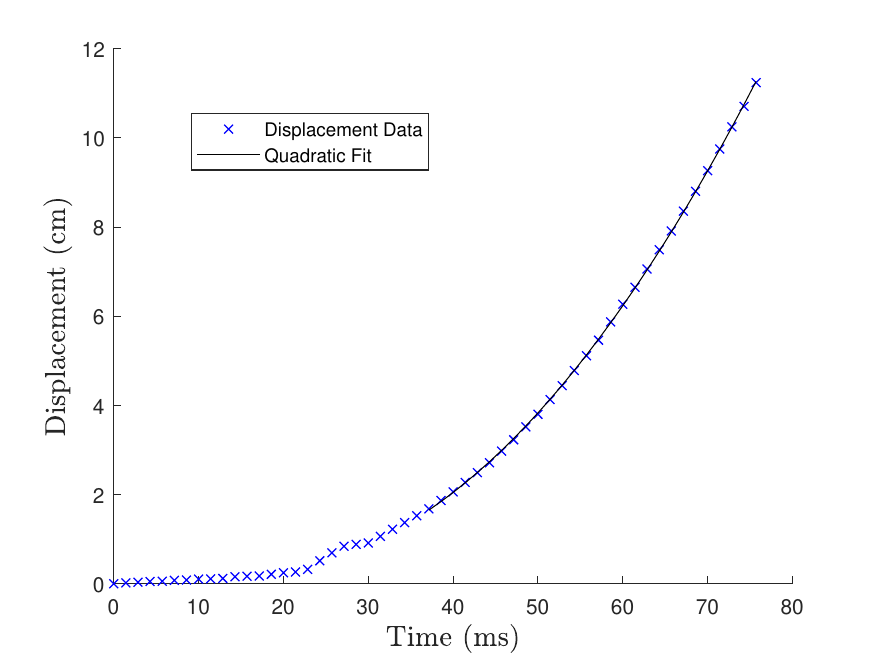}\hspace{5mm}\includegraphics[scale=0.5]{\figdir/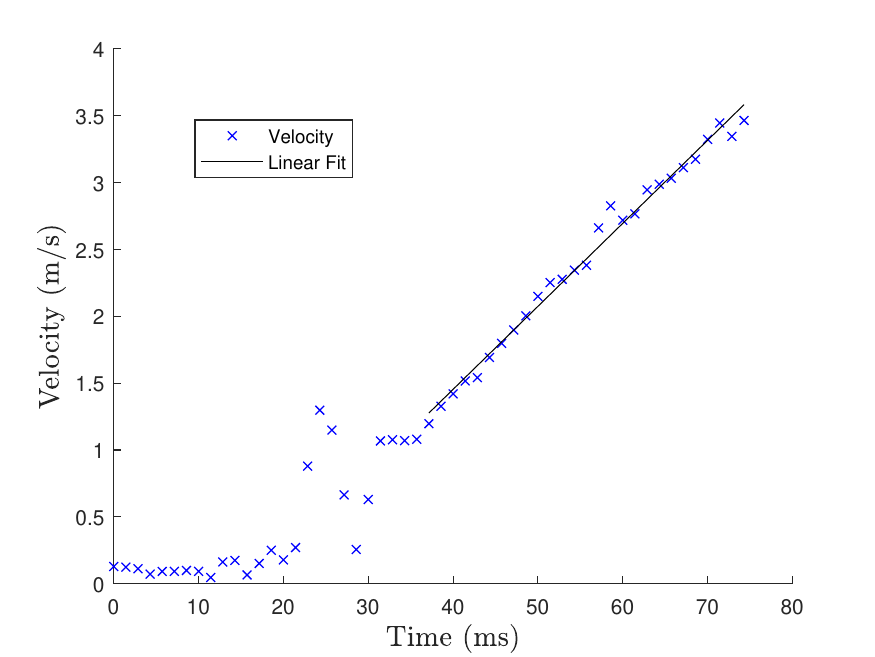} 
 \hspace{85mm}(a)\hspace{85mm}(b)

\caption{Panel (a) shows the spindle displacement as a function of time (ms) along with a quadratic fit made directly to the raw displacement data. Panel (b) shows the velocity of the spindle, as a function of time (ms), and a linear
fit during the last 40 ms.\label{fig:The-Marker-acceleration}}
\end{figure}

The pictures of the wire motion are processed with the free image
analysis software Tracker.\citep{TRACKER} The raw displacement data
from an example run are presented in panel (a) of Fig.~\ref{fig:The-Marker-acceleration}.
The displacement curve has an overall upward curvature indicative
of acceleration. A quadratic fit to these data, $d\left(t\right)=\frac{at^{2}}{2}+v_{0}t+d$,
is plotted with a solid line. 

Note that the 0 second mark on the time axis does not necessarily coincide with
the release time, which is not registered. If the first 29 ms
are excluded, the fit results in the acceleration of $63.7\text{ \ensuremath{\frac{\text{m}}{\text{s}^{2}}}}$.
Assuming gravitational acceleration in Edmonton, Alberta, to be $g=9.838\text{ \ensuremath{\frac{\text{m}}{\text{s}^{2}}}}$,
this translates into about $6.48g$. A fit starting at 35 ms gives
instead $62.1\text{ \ensuremath{\frac{\text{m}}{\text{s}^{2}}}\;}               $
or $6.31g$. 

An alternative analysis is based on the velocity of the spindle, determined
from the displacement data, plotted in Fig.~\ref{fig:The-Marker-acceleration}.
Panel (b) shows the velocity determined from neighboring data points,
based on the smooth central difference formula
$v_{i}=\frac{d_{i+1}-d_{i-1}}{2\Delta t}$. 

The plot clearly illustrates linear velocity growth starting
around 35 ms. A linear fit to the region between 35 and 75 ms is sloped
at about $6.33g$. 

Below 30 ms, the displacement and the velocity data show a feature
that may have resulted from the crude release of the wire (by hand)
and possibly from the rapid initial contraction of the wire. This
region is excluded from the analysis. 

The final value of the spindle acceleration in the 35-75 ms window
is $6.3\left(2\right)g$, where the conservative error estimate is
based on the sensitivity to the starting point of the fit.

\section{Discussion and Summary}

The value of the spindle acceleration, $6.3\left(2\right)g$, is in
reasonable agreement with the rough theoretical prediction for maximum acceleration of about
$10g$ made in Section \ref{sec:Simple-derivation}. That estimate
ignored all dissipative effects, the inertia of the pulley, and even
the weight of the spindle and of the wire. It was based only on the
added mass, which is therefore found to crucially influence the motion
of a body immersed in a fluid. If the added mass were ignored, the
buoyancy force of 4 N would have accelerated a body of mass $10.8$
gram at almost $40g$. 

The experiment described here clearly shows
that the common textbook neglect of the added mass is unrealistic
and misleading, even for a slender body for which the added mass is
relatively small. On the other hand, the experiment shows that the
buoyancy can accelerate a body in fluid by significantly more than
$2g$.

The theoretical value of acceleration may be refined by taking into account the 
finite mass of the styrofoam body. 
Then, Newton's Second Law gives the acceleration
\begin{equation}
a =\frac{m_{b}-m_{f}}{m_{b}+\Cm m_{f}} g,\label{eq:NewtonLaw}
\end{equation}
where $m_{b}$ is the measured ellipsoid mass of 10.8 grams, $m_{f}$ is
the mass of the displaced $0.44$ liter of water, and $\Cm \simeq 0.1$ is
the added mass coefficient  obtained in the slender body approximation.
Including the body mass suppresses the predicted acceleration to $7.3g$, in better 
agreement with the measured value $6.3\left(2\right)g$.

A full analytic solution to the added mass of the prolate spheroid,
see Appendix \ref{sec:Added-mass-ellipsoid}, agrees well with the
simple expression derived in Section \ref{sec:Simple-derivation} even
for $\epsilon\simeq0.3$.  The simple derivation alleviates the
complexity of the complete solution in Appendix
\ref{sec:Added-mass-ellipsoid} without sacrificing significant
accuracy as long as $\epsilon$ is fairly small.

The full analytic expression for the added mass coefficient gives rise
to a theoretical acceleration of $8.3g$, further from the
experimentally measured value than the leading slender body
approximation. Higher order terms in the slenderness parameter reduce
the added mass and thus increase the predicted acceleration.

The experiment described here can be relatively easily reproduced in class or during
a field trip. It can also be extended: by making more marks on the
wire, separated by less than the field of view of the camera, the
motion can be tracked over a longer range. The acceleration should
decrease exponentially due to drag. Such an observation could be used
to determine the drag coefficient.\citep{pantaleone:2011aa} Also
the maximum height reached by the body after jumping above the water
surface can be measured and compared with a prediction based on the
observed terminal velocity. Another related experiment could involve
an inverted pendulum attached at the bottom of the tank and fully
submerged.

An expensive high-speed camera is not necessary for an interesting
experiment. Modern cell phones reach 120 and even 240 frames per second,
a rate at which about 10 data points would have been obtained in the
window 35-75 ms analyzed in this paper. 
A feature of the setup described here is that all measurements are
done above water, so the experiment can be carried out in a swimming
pool or in a lake. 

Despite its simplicity, this experiment introduces students to the
advanced topic of mass renormalization. It also provides an opportunity
to introduce elementary notions of hydrodynamics. Both limiting cases:
of a sphere (slenderness parameter $\epsilon$ equal 1) and of a slender
body ($\epsilon\ll 1$) can be treated with simple mathematical tools.
In the case of the sphere, the calculation is analogous to the electrostatic
problem of a conducting sphere inserted into a uniform electric field.
Solving both problems in parallel illustrates how both Dirichlet and
Neumann boundary conditions are used with the Laplace equation. 

The slender body case described in Section \ref{sec:Simple-derivation}
is related to the idea of a singular perturbation: the added-mass
coefficient is not analytic in the limit of $\epsilon\to0$ and contains
a logarithm. This is one of the simplest physics problems that exhibits
such a feature and can be used as a starting point of a deeper mathematical
discussion.\citep{bender1999methods}

\begin{acknowledgments} 
We thank Professor Bruce R.~Sutherland for letting us use the
Environmental and Industrial Fluid Dynamics Laboratory to carry out
measurements; Professor David W.~Hertzog for suggesting how to
accurately track the spindle motion; Dr.~Isaac Isaac, Lorne Roth, and
Taylor Rogers for lending us a high-speed camera and other equipment;
and Jordan Cameron and Antonio Vinagreiro for  helping to machine the
spindle. A.C.~thanks Professor {\L}ukasz Turski for a stimulating
discussion on the acceleration of spherical balloons.
We thank Thomas Smid for pointing out an error in our estimate of the
initial static stretch of the copper wire. 
This research was supported by the Natural Sciences and Engineering Research Council of Canada.
\end{acknowledgments}

\appendix

\section{Added mass of a sphere \label{sec:Added-mass-sphere}}
In this paper we assume an ideal fluid, neglecting its viscosity and
compressibility. The viscosity could play a role only in the very first
moments of the motion; however, the magnitude of 
the viscous force is negligible in comparison with buoyancy when the velocity is close to zero. 

Compressibility is negligible because the velocity is much smaller
than the speed of sound throughout the observed motion.

The fluid pattern is assumed to be laminar in the determination of the
added mass coefficient. This assumption is justified at the 
beginning of the motion when the velocity is very small. This is
sufficient to predict the initial acceleration. The Reynolds number
${v R \over \nu}$ (with $v$ denoting the velocity of the body,
$R$ its characteristic size, and $\nu \simeq 10^{-6} \,\text{m}^2/\text{s}$ the kinematic
viscosity of water) is estimated to reach $10^4$ around the first
centimeter of motion. In future studies it may be interesting to add
dye to water to probe for the onset of turbulence. 

\subsection{Velocity potential}

Kinetic energy of a fluid disturbed by a moving sphere is computed
here, to explain the method.\citep{Milne-Thomson-Hydr} The flow being
irrotational, the fluid velocity is a gradient of a scalar function,
the velocity potential, $\vv=\vnabla\phi\left(\vr\right)$. For an
incompressible flow, the divergence of the velocity vanishes and $\phi$
satisfies the Laplace equation $\nabla^{2}\phi=0$. Spherical coordinates
are employed, with the origin at the center of the sphere. Because
of axial symmetry, $\phi$ is independent of the azimuth angle $\varphi$
and depends only on $r$ and $\theta$. Thus
\begin{equation}
r\partial_{r}^{2}\left(r\phi\right)+\partial_{\cos\theta}\left(\sin\theta\partial_{\cos\theta}\phi\right)=0.\label{Laplace:spher}
\end{equation}
In the rest frame of the fluid far from the sphere, the sphere is
moving with speed $v$ along the $z$-axis. Denote the outward-drawn
normal to the sphere surface by $\hat{n}$. The projection of the
velocity $v\hat{z}$ of a surface element on that normal is $vn_{z}$
where $n_{z}=\hat{n}\cdot\hat{z}$. The kinematic boundary condition
for the fluid velocity on the sphere surface $\left(r=c\right)$ is
$\vv\cdot\hat{n}=vn_{z}$ (the fluid is moving in the direction perpendicular
to the surface with the same speed as the surface itself is moving
in that direction). Thus the directional derivative of the velocity
potential along the normal is
\begin{equation}
\frac{\partial\phi}{\partial n}=vn_{z}.\label{BC:kinem}
\end{equation}
Since $n_{z}$ is the projection of the normal on the $z$-axis,
$n_{z}\d n=\d z$. Thus $\phi_{s}\left(r,\theta\right)=vz$ satisfies
the boundary condition (\ref{BC:kinem}) as well as the Laplace equation.
This simple dependence only on $z$ is a general result, valid not
only for the sphere; it will be used also for the ellipsoid. 

However, $\phi_{s}$ does not satisfy the correct boundary condition
at infinity, where the potential should be constant to give zero velocity.
A suppression factor $f\left(r\right)$ is needed, independent of
the direction. In order to find it, substitute into (\ref{Laplace:spher})
\begin{equation}
\phi\left(r,\theta\right)=v\phi_{s}\left(r,\theta\right)f\left(r\right),
\end{equation}
Since $\phi_{s}\left(r,\theta\right)=vz=vr\cos\theta$ satisfies the
Laplace equation, Eq.~(\ref{Laplace:spher}) becomes an ordinary
differential equation for the factor $f\left(r\right)$, in which
only its derivatives appear, 
\begin{equation}
rf^{\prime\prime}\left(r\right)+4f^{\prime}\left(r\right)=0\implies f^{\prime}\left(r\right)=\frac{c_{1}}{r^{4}}\implies f\left(r\right)=\frac{c_{2}}{r^{3}}+c_{3},\label{spherical-suppress}
\end{equation}
where $c_{3}$ must vanish for proper behavior at $r\to\infty$. The
complete velocity potential is
\begin{equation}
\phi\left(r,\theta\right)=vr\cos\theta f\left(r\right)=v\frac{c_{2}}{r^{2}}\cos\theta.
\end{equation}
and the overall normalization $c_{2}=-\frac{c^{3}}{2}$ is fixed by
the boundary condition Eq.~(\ref{BC:kinem}) imposed on the surface
of the sphere $r=c$,
\begin{equation}
\phi\left(r,\theta\right)=-v\frac{c^{3}}{2r^{2}}\cos\theta.
\end{equation}

\subsection{Kinetic energy of the fluid\label{subsec:KE-sphere}}

In order to find the added-mass coefficient for the sphere, compute
the kinetic energy of the fluid; Green's theorem converts the integral
over the volume of the fluid into one over its boundary: only the
surface of the sphere contributes,
\begin{align}
T & =\frac{\rho}{2}\int\left(\vnabla\phi\left(\vr\right)\right)^{2}\d^{3}r=-\frac{\rho}{2}\int\phi\left(r=c,\theta\right)\vnabla\phi\cdot\d^{2}\vS=\frac{\rho}{4}v^{2}\underbrace{\int c\cos\theta\d^{2}S}_{V}=\frac{1}{2}\cdot\frac{m_{f}v^{2}}{2},
\end{align}
where $m_{f}=V\rho$ is, as before, the mass of the displaced fluid
and $V$ is the volume of the sphere.\citep{Milne-Thomson-Hydr} In
conclusion, the added-mass coefficient is $\frac{1}{2}$ for a sphere,
in agreement with Ref.~\onlinecite{pantaleone:2011aa}.

\section{Added mass of a spheroid\label{sec:Added-mass-ellipsoid}}

\subsection{Prolate spheroidal coordinates}

Consider a prolate ellipsoid of revolution (a spheroid), with its
long axis on the $z$-axis, and velocity $\vv$ along $z$. Prolate
spheroidal coordinates $\sigma,\tau,\varphi$ are convenient for finding
the velocity potential of the disturbed fluid. They are related to
Cartesian coordinates by
\begin{align}
x & =\sqrt{\left(\sigma^{2}-1\right)\left(1-\tau^{2}\right)}\cos\varphi\\
y & =\sqrt{\left(\sigma^{2}-1\right)\left(1-\tau^{2}\right)}\sin\varphi\\
z & =\sigma\tau.
\end{align}
Surfaces of constant $\sigma$ are spheroids symmetric around the
$z$-axis. $\varphi$ is the usual azimuth angle; surfaces of constant
$\varphi$ are planes containing the $z$-axis. Finally, $\tau$ parametrizes
a family of two-sheet hyperboloids that intersect the spheroids and
the constant-$\varphi$ planes at right angles. In a limit in which
the spheroids become spheres, these hyperboloids become cones parametrized
by a polar angle; in that sense $\tau$ replaces the $\cos\theta$
of spherical coordinates. 

An element of distance squared is expressed as
\begin{equation}
\left(\d s\right)^{2}=h_{\sigma}^{2}\left(\d\sigma\right)^{2}+h_{\tau}^{2}\left(\d\tau\right)^{2}+h_{\varphi}^{2}\left(\d\varphi\right)^{2}
\end{equation}
with 
\begin{equation}
h_{\sigma}=\sqrt{\frac{\sigma^{2}-\tau^{2}}{\sigma^{2}-1}},\quad h_{\tau}=\sqrt{\frac{\sigma^{2}-\tau^{2}}{1-\tau^{2}}},\quad h_{\varphi}=\sqrt{\left(\sigma^{2}-1\right)\left(1-\tau^{2}\right)}.
\end{equation}

\subsection{Velocity potential}

For an axially-symmetric problem, the velocity potential $\phi\left(\sigma,\tau\right)$
is independent of $\varphi$ and its Laplace equation becomes
\begin{equation}
\partial_{\sigma}\left[\left(\sigma^{2}-1\right)\partial_{\sigma}\phi\right]+\partial_{\tau}\left[\left(1-\tau^{2}\right)\partial_{\tau}\phi\right]=0.\label{Laplace:spheroid}
\end{equation}
On the surface of a spheroid with the ratio of short-to-long axes
$\epsilon=\frac{b}{c}$, the coordinate $\sigma$ is constant, 
\begin{equation}
\sigma=\sigma_{0}=\frac{1}{\sqrt{1-\epsilon^{2}}}.\label{sigma0}
\end{equation}
The boundary condition (\ref{BC:kinem}) on the spheroid surface can
be written as
\begin{equation}
\frac{\partial\phi}{\partial n}=vn_{z}\implies\frac{1}{h_{\sigma}}\frac{\partial\phi}{\partial\sigma}=v\cos\theta_{z}=\frac{v}{h_{\sigma}}\frac{\partial z}{\partial\sigma}\implies\frac{\partial\phi}{\partial\sigma}=v\frac{\partial z}{\partial\sigma}.\label{BC:ellip}
\end{equation}

This condition is again satisfied by $\phi_{s}=vz=v\sigma\tau$ on
the spheroid. All that is needed to complete the solution $\phi\left(\sigma,\tau\right)$
is the suppression factor $f\left(\sigma\right)$. Just like in Appendix
\ref{sec:Added-mass-sphere}, the substitution $\phi=\phi_{s}f$ converts
the Laplace equation (\ref{Laplace:spheroid}) into a first order
ODE for $f^{\prime}\left(\sigma\right)$,
\begin{equation}
\left(\sigma^{2}-1\right)\sigma f^{\prime\prime}\left(\sigma\right)+2\left(2\sigma^{2}-1\right)f^{\prime}\left(\sigma\right)=0.
\end{equation}
Two integrations, with the condition of vanishing $f\left(\sigma\to\infty\right)$,
give
\begin{equation}
f\left(\sigma\right)=C\left(\frac{1}{2}\ln\frac{\sigma+1}{\sigma-1}-\frac{1}{\sigma}\right).
\end{equation}
Far from the origin, this is indeed a suppression factor, behaving
like $f\left(\sigma\to\infty\right)\sim\frac{C}{3\sigma^{3}}$, like
the analogous factor in spherical coordinates, eq.~(\ref{spherical-suppress}).
In order to determine $C$, consider again the boundary condition
on the spheroid, 
\begin{equation}
\partial_{\sigma}\left[v\sigma\tau f\left(\sigma\right)\right]=v\frac{\partial z}{\partial\sigma}=v\tau,\qquad\left(\sigma=\sigma_{s}\right)
\end{equation}
from which follows
\begin{equation}
\frac{1}{C}=f\left(\sigma_{0}\right)+\sigma_{0}f^{\prime}\left(\sigma_{0}\right)=\frac{1}{2}\ln\frac{\sigma_{0}+1}{\sigma_{0}-1}+\frac{\sigma_{0}}{1-\sigma_{0}^{2}},
\end{equation}
and the complete potential reads
\begin{equation}
\phi\left(\sigma,\tau\right)=v\sigma\tau\frac{\frac{1}{2}\ln\frac{\sigma+1}{\sigma-1}-\frac{1}{\sigma}}{\frac{1}{2}\ln\frac{\sigma_{0}+1}{\sigma_{0}-1}+\frac{\sigma_{0}}{1-\sigma_{0}^{2}}}.\label{phi:spheroid}
\end{equation}

\subsection{Added-mass coefficient for a spheroid}

It is convenient to write the potential on the sphere as $\phi\left(\sigma_{0},\tau\right)=vzF$,
where $F=1+\frac{2}{\sigma_{0}\left(\sigma_{0}^{2}-1\right)\ln\frac{\sigma_{0}+1}{\sigma_{0}-1}-2\sigma_{0}^{2}}$.
In analogy with Appendix \ref{subsec:KE-sphere}, the integration
in the kinetic energy expression gives the volume of the spheroid,
and
\begin{align}
T & =-F\frac{m_{f}v^{2}}{2}.
\end{align}
Thus the added-mass coefficient is just $\Cm=-F$. In terms of the
slenderness parameter $\epsilon$, using Eq.~(\ref{sigma0}),
\begin{equation}
\Cm=\frac{1-\epsilon^{2}}{1-\frac{\epsilon^{2}}{\sqrt{1-\epsilon^{2}}}\ln\frac{1+\sqrt{1-\epsilon^{2}}}{\epsilon}}-1.
\end{equation}
This general result has correct limiting behaviors: For a sphere,
$\epsilon=1$, it correctly reproduces $\Cm\left(\epsilon=1\right)=\frac{1}{2}$,
derived in Appendix \ref{subsec:KE-sphere}. For a slender spheroid,
$\epsilon\to0$, 
\begin{equation}
\Cm\left(\epsilon\to0\right)\sim\epsilon^{2}\ln\frac{1}{\epsilon},
\end{equation}
in agreement with the derivation of Section \ref{sec:Simple-derivation}.


\end{document}